\documentclass[12pt]{article}
\usepackage{graphicx}
\usepackage{amssymb}
\usepackage{amscd}
\usepackage{amsmath}
\usepackage{appendix}
\begin{document}
\begin{titlepage}

%\begin{center}
%{\hbox to\hsize{
%\hfill \bf hep-ph/??? }}
{\hbox to\hsize{\hfill March 2015 }}

\bigskip \vspace{3\baselineskip}

\begin{center}
{\bf \large 

Generating Luminous and Dark Matter During Inflation }

\bigskip

\bigskip

{\bf Neil D. Barrie and Archil Kobakhidze \\ }

\smallskip

{ \small \it
ARC Centre of Excellence for Particle Physics at the Terascale, \\
School of Physics, The University of Sydney, NSW 2006, Australia \\
E-mails: nbar5465@physics.usyd.edu.au, archilk@physics.usyd.edu.au 
\\}

\bigskip
 
\bigskip

\bigskip

{\large \bf Abstract}

\end{center}
\noindent 
We propose a new mechanism for generating both luminous and dark matter during cosmic inflation. According to this mechanism, ordinary and dark matter carry common charge which is associated with an anomalous $ U(1)_{X} $ group. Anomaly terms source $ \mathcal{CP} $ and $ U(1)_{X} $ charge violating processes during inflation, producing corresponding non-zero Chern-Simons numbers which are subsequently reprocessed into baryon and dark matter densities. The general framework developed is then applied to two possible extensions of the Standard Model with anomalous gauged $B$ and $B-L$, each with an additional dark matter candidate.
 
\end{titlepage}

\vspace{1cm}

\section{Introduction}

Compelling evidence for cosmic inflation \cite{Guth:1980zm} has been accumulated in a number of astrophysical observations \cite{Ade:2013ktc, Ade:2015tva}. Inflation not only solves several problems of the standard Big Bang theory, but provides an elegant mechanism for generation of primordial inhomogeneities through the quantum fluctuations of an inflaton field \cite{Mukhanov:1981xt}. These spatial irregularities  ultimately led to the formation of galaxies and are imprinted in the temperature anisotropies of the cosmic microwave background radiation. It has been suggested that inflation may play an even more prominent role by also generating the observed matter-antimatter asymmetry in the universe \cite{Barrie:2014waa, Alexander:2004us}. Namely, in \cite{Barrie:2014waa} we argued that a successful baryogenesis scenario can be realised during inflation within models containing anomalous gauge symmetries \cite{Preskill}. According to this mechanism the anomaly term acts as a $ \mathcal{CP} $ and baryon number violating source for production of non-zero Chern-Simons number carried by the electroweak gauge boson, which subsequently generates non-zero baryon number via the anomaly.

In this paper we apply the mechanism proposed in \cite{Barrie:2014waa}  to the simultaneous generation of luminous and dark matter. The idea of a common origin of luminous and dark matter traces back to 90's \cite{Dodelson:1989ii} and has received a renewed interest in recent years (see a review in \cite{Petraki:2013wwa} and references therein). The major motivation to this hypothesis comes from the observation that the present-day mass density of dark matter is about a factor of five higher than the density of visible matter \cite{Ade:2013ktc},
\begin{equation}
\rho_{\rm DM}\simeq 5.5\rho_B~.
\label{1}
\end{equation}
The similarity in these observed densities perhaps indicates towards a strong connection between the physics and cosmological evolution of visible and dark matter. The present-day density, $\rho_B$, of visible matter is believed to be  due to the  tiny excess of baryons over antibaryons generated in the early universe, which is quantified by the baryon asymmetry parameter:
\begin{equation}
\eta_B=\frac{n_b-n_{\bar b}}{s}\simeq 8.5 \cdot 10^{-11}~, 
\label{2}
\end{equation}
where $n_b$ ($n_{\bar b}$) is the baryon (antibaryon) number density and $s$ is the entropy density. Hence, within this picture a similar asymmetry is expected to be generated among dark matter particles and antiparticles.  In our model, visible and dark matter are connected by a common anomalous gauged $U(1)_{X}$, which we introduce in addition to the gauge group of the Standard Model. Then simultaneous generation of luminous and dark matter proceeds during inflation as suggested in Ref. \cite{Barrie:2014waa}. For other recent works relating the ordinary matter-antimatter asymmetry with the asymmetry in the dark sector within gauged $U(1)_X$ extensions of the Standard Model see \cite{Dulaney:2010dj}.       

The rest of the paper is structured as follows. In section \ref{Sec2} we describe a generalised form of the model and the two specific examples we will consider. Section \ref{Sec3} considers the dynamics of the anomalous gauge field during the inflationary epoch. In section \ref{Sec4} we calculate asymmetries generated during inflation in the visible and dark sectors of the theory, and discuss their subsequent evolution in section \ref{Sec5}. The last section is reserved for conclusions. An appendix has been included to provide further details on our calculations. 

\section{Models with an anomalous $U(1)_{X}$ \label{Sec2}}

We consider an extension of the Standard Model that is based on $SU(3)\times SU(2)\times U(1)_{Y}\times U(1)_{X}$ gauge group and contains an additional fermion/s that shall act as a dark matter candidate. The introduction of a scalar to play the role of the inflaton is also required, but the detailed dynamics of inflation is not important for our analysis. The new $ U(1)_{X} $ gauge  symmetry is assumed to be anomalous, and hence the corresponding gauge boson will be necessarily massive with gauge invariance realised non-linearly. The longitudinal degree of freedom of this $ U(1)_{X} $ gauge field is then described by a scalar field $ \theta(x) $, which allows anomaly cancellation through the introduction of appropriate counter terms \cite{Preskill}.\footnote{This theory can be viewed as a low energy limit of an anomaly-free theory, either within ordinary QFT or string theory.} In the presence of a cubic anomaly $ U(1)_{X}^{3} $, the additional Lagrangian terms important to our analysis are given below,
\begin{align}
\frac{1}{\sqrt{-g}}{\cal L}_X=& -\frac{1}{4}g^{\mu\alpha}g^{\nu\beta}X_{\mu\nu}X_{\alpha\beta}+\frac{1}{2}f_X^2g^{\mu\nu}\left(g_X X_{\mu}-\partial_{\mu}\theta \right)\left(g_X X_{\nu}-\partial_{\nu}\theta \right) \nonumber 
\\ &
-{\cal A}_{2}\frac{g_X^2}{16\pi^2}\theta(x)X_{\mu\nu}\tilde X^{\mu\nu}~,
\label{L}
\end{align}    
 where $X_{\mu\nu}$ denotes the field strength of the $U(1)_{X}$ gauge boson with corresponding coupling constant $g_X= m_{X}/f_{X}$, $f_X$ is a parameter that defines the mass of the $U(1)_{X}$ boson ($ m_X $), and $\tilde{X}^{\mu\nu}=\frac{1}{2\sqrt{-g}}\epsilon^{\mu\nu\rho\sigma}X_{\rho\sigma }$ is the dual field strength, in which $\epsilon^{\mu\nu\rho\sigma}$ is the Levi-Civita tensor. We have omitted fermion interactions and the charged current $ j_{X} $ terms. The final term in Eq. (\ref{L}) is responsible for maintaining gauge invariance of the full quantum theory description under $U(1)_{X}$ transformations, and is reminiscent of the term for a massive gauge boson in the Stuekelberg formalism \cite{Stueckelberg:1900zz}. In the proceeding analysis any associated gravitational anomaly is ignored as it is considered to be negligible with respect to the other anomalous contributions. 

In this paper we shall consider two applications of this general model for cogenesis. Namely, we consider the extensions involving an anomalous gauged $ B-L $ and $ B $.  These models contain a fermionic field(s) $ \psi $, which carries a chiral charge under the anomalous gauge symmetry and is sterile under the Standard Model gauge symmetry. The charges of each of the fermions under these additional gauge symmetries are given in Table \ref{smblt1}. The mass $ m_{\psi} $ is an extra parameter which can be directly introduced within the non-linear realisation of the anomalous gauge symmetry. Typically, such a mass is also generated radiatively within the low-energy effective theory, reflecting a more conventional mechanism for mass generation within an ultraviolet anomaly-free completion \cite{prep}.

\begin{table}[ht]
\centering
\begin{tabular}{|c|c|c|c|c|c|}
\hline
Fermion Field & $ SU(3) $ & $ SU(2) $ & $ U(1)_{Y} $& Case 1 $ U(1)_{B-L} $  & Case 2 $ U(1)_{B} $ \\ \hline
 $Q^{i}_{L} = \begin{pmatrix}u\\d\end{pmatrix}_{L}^{i}$ & $ 3 $ & $ 2 $ & $ 1/6 $ & $ 1/3 $ & $ 1/3 $ \\ \hline
$u^{i}_{R}$ & $ 3 $ & $ 1 $ & $ 2/3 $ & $ 1/3 $ & $ 1/3 $ \\ \hline
$d^{i}_{R}$ & $ 3 $ & $ 1 $ & $ -1/3 $ & $ 1/3 $ & $ 1/3 $ \\ \hline
$L^{i}$ = $\begin{pmatrix}\nu\\e\end{pmatrix}_{L}^{i}$ & $ 1 $ & $ 2 $ & $ -1/2 $ & $ -1 $ & $ 0 $ \\ \hline
$e^{i}_{R}$ & $ 1 $ & $ 1 $ & $ -1 $ & $ -1 $ & $ 0 $ \\ \hline
$\psi$ & $ 1 $ & $ 1 $ & $ 0 $ & $ q_{\psi} $ & $ q_{\psi} $ \\ \hline
\end{tabular}
\caption{\small{The representations of the Standard Model and dark fermion $\psi$ in reference to the gauge symmetries.}}
\label{smblt1}

\end{table}
\subsection*{Case 1: $U(1)_{B-L}$ and a sterile fermion $ \psi $}
In the Standard Model an additional $U(1)_{B-L}$ gauge symmetry is anomalous unless three right handed neutrinos are introduced. The associated anomalies are trace and cubic: ${\cal A}_{0}(U(1)_{B-L})=-3$ and ${\cal A}_{1}(U(1)_{B-L}^{3})=-3$. We introduce $N_{\psi}$ new right-handed (for definiteness) Weyl fermions $ \psi$, some of which act as dark matter candidates in our model. For simplicity we assume that they carry the same $B-L$ charge $q_{\psi}$ and interact only via exchange of the $ B-L $ gauge boson. The addition of these states alters the $B-L$ anomalies as follows: ${\cal A}_{0}:={\cal A}_{0}(U(1)_{B-L})=-N_{\psi}q_{\psi}-3$ and ${\cal A}_{1}:={\cal A}_{1}(U(1)_{B-L}^{3})=-N_{\psi}q_{\psi}^3-3$. In this case the dark matter fermion does not introduce any new anomalies. We will ignore the gravitational anomaly $ \mathcal{A}_{0} $ in our analysis, but it should be noted that taking $ q_{\psi}=-3 $ or $ -1 $ and $ N_{\psi}=1 $ or $ 3 $, respectively,  eliminates this gravitational anomaly leaving only the cubic anomaly non-zero. Obviously, $N_{\psi}=3$, $q_{\psi}=-1$ removes all anomalies, so we are not interested in such charge assignment in this paper. 

The addition of the $ B-L $ gauge symmetry and dark matter candidate to the Standard Model leads to a Lagrangian density of the same form given for the general case presented in Eq. (\ref{L}).

\subsection*{Case 2: $U(1)_{B}$ and a sterile baryon $ \psi $}

Gauging the baryon number symmetry of the Standard Model results in the inclusion of two mixed anomalies involving the weak and hypercharge gauge groups: 
${\cal A}_{2}(SU(2)^2\times U(1)_{B})=3/2$ and ${\cal A}_{3}(U(1)_Y^2\times U(1)_{B})=-3/2$.
The addition of a new sterile state $ \psi $ leaves these mixed anomalies unchanged, but introduces two new unmixed anomalies: ${\cal A}_{0}:={\cal A}_{0}(U(1)_{B})=-N_{\psi}q_{\psi}$ and ${\cal A}_{1}:={\cal A}_{1}(U(1)_{B}^{3})=-N_{\psi}q_{\psi}^3$.
Hence there are now four anomalies, each of which will contribute to baryonic charge generation during the inflationary epoch, but only two of which include generation of fermions in the dark matter sector (${\cal A}_{0}$ and ${\cal A}_{1}$).

The presence of additional mixed anomalies means that extra anomaly cancelling terms are required with respect to the gauged $ B-L $ case considered above, i.e.
\begin{align}
\frac{1}{\sqrt{-g}}{\cal L}_X=& -\frac{1}{4}g^{\mu\alpha}g^{\nu\beta}X_{\mu\nu}X_{\alpha\beta}+\frac{1}{2}f_X^2g^{\mu\nu}\left(g_X X_{\mu}-\partial_{\mu}\theta \right)\left(g_X X_{\nu}-\partial_{\nu}\theta \right) \nonumber 
\\ &
-{\cal A}_{1}\frac{g_X^2\theta(x)}{16\pi^2}X_{\mu\nu}\tilde X^{\mu\nu}-{\cal A}_{2}\frac{g_1^2\theta(x)}{16\pi^2}B_{\mu\nu}\tilde B^{\mu\nu}
-{\cal A}_{3}\frac{g_2^2\theta(x)}{16\pi^2}W^{a}_{\mu\nu}\tilde W^{a\mu\nu}
\label{L1}
\end{align}    
 where $B_{\mu\nu}$ and $W_{\mu\nu}$ denote the hypercharge and weak field strengths respectively, with corresponding coupling constants $ g_{1} $ and $ g_{2} $.

\section{Dynamics of an anomalous gauge field during inflation}
\label{Sec3}
For a model to successfully produce a charge asymmetry in the early universe it must satisfy the well known Sakharov conditions \cite{Sakharov:1967dj}. We will now discuss the framework of our new mechanism for cogenesis and how it satisfies these criteria.

 Firstly, we wish to describe the universe using the Robertson-Walker metric tensor, which represents a homogeneous, isotropic and spatially flat cosmological spacetime. In conformal coordinates the metric can be expressed as: $g_{\mu\nu}=a^2(\tau)\eta_{\mu\nu}$. During inflation the scale factor $a(\tau)$ is given by the following:
 \begin{equation}
 a(\tau)=-1/H_{\rm inf}\tau~,
 \label{a}
 \end{equation} 
 where $H_{\rm inf}$ is the expansion rate during inflation ($H_{\rm inf}\cong {\rm const.}$) and $\tau \in [-\infty, 0]$ is the conformal time.  
 
% We assume that $g_X\ll 1$,          
 To allow analytical treatment, the analysis that follows requires certain simplifying assumptions. For the $\theta$ field we only consider a classical homogeneous background configuration, $\theta(\tau, \vec{x})=\theta(\tau)$, and ignore quantum fluctuations over it. We take $g_{X}\ll 1$ such that the $\theta (x)$ and $X_{\mu}(x)$ fields essentially decouple from each other. This also implies that the $U(1)_{X}$ boson is light relative to the scale $ f_{X} $, $m_X/f_X\ll 1$, and hence we will not be interested in its dynamics during inflation. With these assumptions the Lagrangian Eq. (\ref{L1}) becomes:
 \begin{align}
 {\cal L}_X =& -\frac{1}{4}\eta^{\mu\alpha}\eta^{\nu\beta}X_{\mu\nu}X_{\alpha\beta}+\frac{1}{2}a(\tau)^2\eta^{\mu\nu}\left(m_X X_{\mu}-\partial_{\mu}\phi(\tau) \right)\left(m_X X_{\nu}-\partial_{\nu}\phi(\tau) \right) \nonumber 
  \\ &
 -{\cal A}_{1}\frac{g_X^2\phi(\tau)}{32\pi^2f_{X}} \epsilon^{\mu\nu\alpha\beta} X_{\mu\nu}X_{\alpha\beta}~,
 \label{L2}
 \end{align}    
where $\phi(\tau)\equiv f_X\theta(\tau)$.   
From this Lagrangian follows the equation of motion for $\phi(\tau)$:
\begin{equation}
\left(a^2\phi'\right)'=0~,
\label{6}
\end{equation}
where  $\phi'\equiv d\phi/d\tau$ and we have ignored any terms quadratic in $X_{\mu}$. Solving for $\phi'(\tau)$ we obtain:
\begin{equation}
\phi'(\tau)=\frac{\phi'_0}{a^2(\tau)}~,
\label{7}
\end{equation}
where $\phi'_0$ is an integration constant associated with the `field velocity' at the start of inflation, which is defined at $\tau=\tau_0$, where $ a(\tau_0)=1 $. Substituting Eq. (\ref{7}) into the linearized equation of motion for the $X_{\mu}$ gauge field gives:

\begin{equation}
\left(\partial_{\tau}^2-\vec \bigtriangledown^2 + \left(\frac{m_{X}}{H_{\rm inf}\tau}\right)^2 \right) X^{i}+\kappa_{X} \tau^2 \epsilon^{ijk}\partial_jX_k=0~,
\label{xfieldeom}
\end{equation}
where \begin{equation}  \kappa_{X} =|{\cal A}_{1}|\frac{g_X^2\phi'_0H_{\rm inf}^2}{4\pi^2 f_X}~.
\end{equation} 
and the gauge $X_0=\partial_i X_i=0$ has been chosen. The source of ${\cal CP}$ violation in our model is apparent in Eq. (\ref{xfieldeom}) where the two terms have opposite ${\cal P}$, and hence, ${\cal CP}$ transformations.   

In the discussion that follows we treat the $U(1)_X$ gauge boson as a massless particle, as we have assumed $m_X\ll H_{\rm inf}$. To then quantize this model we promote the $ X $ gauge boson fields to operators and assume that the boson has two possible circular polarisation states:
\begin{equation}
X_i=\int\frac{d^3\vec k}{(2\pi)^{3/2}}\sum_{\alpha}\left[G_{\alpha}(\tau,k)\epsilon_{i\alpha}\hat a_{\alpha} {\rm e}^{i\vec k\cdot\vec x}+
G^{*}_{\alpha}(\tau,k)\epsilon^{*}_{i\alpha}\hat a_{\alpha}^{\dagger}{\rm e}^{-i\vec k\cdot\vec x}
\right]~,
\label{11}
\end{equation}
where $\vec \epsilon_{\pm}$ denotes the two possible helicity states of the $U(1)_{X}$ gauge boson ($\vec \epsilon_{+}^{*}=\vec \epsilon_{-}$) and the creation, $\hat a_{\alpha}^{\dagger}(\vec k)$, and annihilation, $\hat a_{\alpha}(\vec k)$, operators satisfy the canonical commutation relations:
\begin{equation}
\left[\hat a_{\alpha}(\vec k), \hat a_{\beta}^{\dagger}(\vec k')\right]=\delta_{\alpha\beta}\delta^3(\vec k-\vec k')~.
\end{equation}
and      
\begin{equation}
\hat a^a_{\alpha}(\vec k)\vert 0\rangle_{\tau}=0~. 
\label{14}
\end{equation}
where $\vert 0\rangle_{\tau}$ is an instantaneous vacuum state at time $\tau$.
 
 The mode functions in Eq. (\ref{11}) are described by the following equations, from Eq. (\ref{xfieldeom}),
\begin{equation}
G''_{\pm}+\left(k^2 + \frac{\lambda^2}{\tau^2}\mp \kappa_{X}\tau^2 k \right)G_{\pm}=0~, 
\label{modefunc}
\end{equation}  
where  $ \lambda=\frac{m_{X}}{H_{\rm inf}} $, which is assumed to be small as stated above.

The equations for the mode functions $ G_{\pm} $ given in Eq. (\ref{modefunc}) have the following solutions,

\begin{equation}
G_{+}(\tau,k)= 2^{\frac{1+\nu}{2}} e^{-z}{2} \tau^{\frac{1}{2}+\nu} \left[ C_{1} U\left(\frac{1+\nu}{2}-\frac{\Omega_{k}}{4},1+\nu,z\right)
+ C_{2} M\left(\frac{1+\nu}{2}-\frac{\Omega_{k}}{4},1+\nu,z\right)
\right]
\label{G+}
\end{equation}
and 
\begin{equation}
G_{-}(\tau,k)= 2^{\frac{1+\nu}{2}} e^{z}\tau^{\frac{1}{2}+\nu}
\left[C_3 U\left(\frac{1+\nu}{2}-\frac{i\Omega_{k}}{4},1+\nu,\frac{z}{i}\right)  +C_4 M\left(\frac{1+\nu}{2}-\frac{i\Omega_{k}}{4},1+\nu,\frac{z}{i}\right)\right]
\label{G-}
\end{equation}
where $ z=\frac{k^2 \tau^2}{\Omega_{k}} $, $\Omega_k=\left(\frac{k^3}{\kappa_{X}}\right)^{1/2}$, $\nu=\frac{1}{2}\sqrt{1-4\lambda^2}\sim\frac{1}{2}-\lambda^2$, $ U(a,b,z) $ is a confluent hypergeometric function of the second kind, and $ M(a,b,z) $ is a confluent hypergeometric function of the first kind (Kummer Function).

In the limit $  |\tau|\rightarrow 0$ (or $ k^2 + \frac{\lambda^2}{\tau^2}\gg \kappa_{X}\tau^2 k $), $\mathcal{CP}$-invariant wave modes are obtained. These are described by,
\begin{equation}
X_i=\int\frac{d^3\vec k}{(2\pi)^{3/2}}\sum_{\alpha}\left[F_{\alpha}(\tau,k)\epsilon_{i\alpha}\hat b_{\alpha} {\rm e}^{i\vec k\cdot\vec x}+
F^{*}_{\alpha}(\tau,k)\epsilon^{*}_{i\alpha}\hat b_{\alpha}^{\dagger}{\rm e}^{-i\vec k\cdot\vec x}
\right]~,
\label{11a}
\end{equation}
where the wave mode functions $F_{\pm}$ are found to be,
\begin{equation}
F_{+}(\tau,k)=\frac{\sqrt{\pi \tau}}{2} H^{(2)}_{\nu}(k\tau) e^{-i\frac{\pi}{2}(\frac{1}{2}+\nu)}
~~\textrm{and}~~
F_{-}(\tau,k)=\frac{\sqrt{\pi \tau}}{2} H^{(1)}_{\nu}(k\tau) e^{i\frac{\pi}{2}(\frac{1}{2}+\nu)}
\label{16}
\end{equation}

By matching the modes in Eq. (\ref{G+}) and (\ref{G-}) to those in Eq. (\ref{16}) and using the known Wronskian normalisation we can determine the coefficients $ C_{1-4} $. For more details on this calculation and the form of the coefficients see App. \ref{A}.

We can now compare the birefringent and $\mathcal{CP}$-invariant modes to derive the Bogoluibov coefficients relating the two sets of creation and annihilation operators, $\lbrace \hat a_{\alpha}^a, \hat a^{a\dagger}_{\alpha} \rbrace$ and $\lbrace\hat b^a_{\alpha}, \hat b^{a\dagger}_{\alpha} \rbrace$, in Eqs ($ \ref{11} $) and ($ \ref{11a} $). The Bogoluibov transformations are defined by,
\begin{eqnarray}
\label{bog1}
\hat b^a_{\alpha}(\vec k)= \alpha_{\alpha} a^{a\dagger}_{\alpha}(\vec k)+\beta^{*}_{\alpha}\hat a_{\alpha}^{a}(\vec k) \\
\label{bog2}
\hat b^{a\dagger}_{\alpha}(\vec k)= \alpha^{*}_{\alpha} a^{a}_{\alpha}(\vec k)+\beta_{\alpha}\hat a_{\alpha}^{a\dagger}(\vec k)
\end{eqnarray}

In this scenario the relevant Bogoliubov coefficients are found to be:
\begin{equation}
\alpha_{\pm}=1-\frac{1}{2^{1-\nu}}\left(1\pm \frac{i \lambda^2}{(k\tau)^{1-2\lambda^2}}\left(1-\frac{\pi(k\tau)^{1-2\lambda^2} }{2^{\nu}} \right) \mp \frac{i2^{1-\nu}(k\tau)^{\lambda^2}}{\sqrt{k}}e^{\mp i \pi \lambda^2 /2} G^{\prime *}_{\pm}|_{\frac{\kappa\tau^2}{k},k|\tau|\to 0}\right)
\end{equation}

\begin{equation}
\beta_{\pm }=\frac{e^{\mp i \pi \lambda^2}}{2^{1-\nu}}\left(1\mp \frac{i \lambda^2}{(k\tau)^{1-2\lambda^2}}\left(1-\frac{\pi(k\tau)^{1-2\lambda^2} }{2^{\nu}} \right) \pm \frac{i2^{1-\nu}(k\tau)^{\lambda^2}}{\sqrt{k}}e^{\pm i \pi \lambda^2 /2} G^{\prime *}_{\mp}|_{\frac{\kappa\tau^2}{k},k|\tau|\to 0}\right)
\end{equation}
where we have considered the superhorizon modes ($k|\tau|\approx 0$).

\section{Simultaneous generation of luminous and dark matter during inflation \label{Sec4}}

We can now calculate the general $ X $ charge density generated during inflation. It is known that the anomalous non-conservation of the $ X $ charge current is given by,
\begin{equation}
\partial_{\mu}\left(\sqrt{-g} j_X^{\mu}\right)={\cal A}_{1}\frac{g_X^2}{32\pi^2}\epsilon^{\mu\nu\rho\sigma} 
X_{\mu\nu} X_{\rho\sigma}\equiv {\cal A}_{1}\frac{g_X^2}{8\pi^2}\partial_{\mu}\left(\sqrt{-g} K^{\mu}\right)~,
\label{18}
\end{equation}
where $K^{\mu}=\frac{1}{2\sqrt{-g}}\epsilon^{\mu\nu\rho\sigma}X_{\nu\rho}X_{\sigma}$ is a topological current. This implies that the net $ X $ charge density $n_X=n_{\rm x}-n_{\bar {\rm x}}\equiv a^{-1}(\tau)\langle 0 \vert j_X^0\vert 0 \rangle$ is related to the Chern-Simons number density of the $U(1)_{X}$ gauge boson by the following equation,
\begin{eqnarray}
n_X= |{\cal A}_{1}|\frac{g_X^2}{8\pi^2} a(\tau_{\rm end}) n_{CS}~,
\label{19}
\end{eqnarray}
where $\tau=\tau_{\rm end}$ is the conformal time at the end of inflation, and $n_X(\tau_0)= n_{CS}(\tau_0)=0$ at the start of inflation.
The form of the Chern-Simons number is given below, in which we wish to consider only large scale superhorizon modes ($ k|\tau|\simeq0 $).
\begin{align}
\begin{split}
n_{CS}&=  \frac{1}{a^4(\tau_{\rm end})}\epsilon^{ijk}\lim_{k|\tau|\to 0}\langle 0\vert X_i \partial_j X_k\vert 0\rangle \\ & 
\simeq \frac{1}{4\pi^2 a^4(\tau_{\rm end})}\int_{\mu}^{\Lambda}k dk\left[\left\vert G^{\prime}_{+}\right \vert^2_{\frac{\kappa\tau^2}{k},k|\tau|\to 0}-\left \vert G^{\prime}_{-}\right \vert^2_{\frac{\kappa\tau^2}{k},k|\tau|\to 0}\right]-{\cal O}(\lambda^2)~,
\end{split} 
\label{20}
\end{align}
where we ignore small terms  with quadratic or higher orders of $\lambda$.  Note that the upper limit in the integral in Eq. (\ref{20}) simply cuts out sub-horizon modes for which our approximate calculations are not applicable. The ultraviolet modes do not give significant contribution anyway, since they act as $ \mathcal{CP} $-invariant planewaves which expectantly leads to a cancellation between the positive and negative frequency modes. The dominant contribution to $n_{CS}$  is given by infrared modes, and in fact the integral is divergent. This divergence is a reminiscent of the well-known infrared divergence of de Sitter-invariant two-point functions, which possibly signals that pure de Sitter approximation of inflationary phase becomes inadequate in our case. There is no commonly accepted prescription for regularization of this type of divergences in the literature and we simply introduce an infrared cut-off $\mu$.  

We assume that the only non-negligible source of entropy density is reheating after inflation, for which the entropy density produced is: $s \simeq \frac{2\pi^2}{45}g_{*}T^3_{\rm rh}$, where $T_{\rm rh}$ is the associated reheating temperature and $ g_{*}(T_{\rm rh})\sim 100 $. Upon taking a first order expansion around $ \Omega_{k}=0 $ in Eq. (\ref{20}) we obtain the following expression for the $ X $ charge asymmetry parameter generated by the unmixed anomaly:
\begin{align}
\begin{split}
\eta_X=& \frac{n_X}{s}\approx   |{\cal A}_{1}| \frac{30 g_{X}^{2}}{\pi^{10}g_{*}} \Gamma\left(\frac{3}{4}\right)^4 e^{-3 N_{\rm e}}\left(\frac{\kappa_X}{ \mu T_{\rm rh}^2}\right)^\frac{3}{2}\\ & \approx   7\cdot 10^{-11} |{\cal A}_{1}|^{5/2}  \left(\frac{m_{X}}{10^{12}~{\rm GeV}}\right)^{5} \left(\frac{\phi'_0}{M_{p}^{2}}\right)^{\frac{3}{2}} \left(\frac{H}{10^{14}~{\rm GeV}}\right)\left(\frac{T_{\rm rh}}{10^{16}~{\rm GeV}}\right)^{-2} \\ & \times \left(\frac{f_X}{10^{14}~{\rm GeV}}\right)^{-\frac{13}{2}} \left(\frac{\mu}{  10^{-42}~{\rm GeV}}\right)^{-\frac{3}{2}}~,
\end{split}
\label{24}
\end{align} 
where $ N_e $ denotes the minimum number of e-folds required to solve the horizon and flatness problems, and includes the additional dilution that occurs if the reheating period is not instantaneous. The number of e-folds that contribute to the dilution of $ n_{X} $ is,
\begin{equation}
N_e= N_{\rm inf} + N_{\rm rh} \simeq 27.5 + \frac{2}{3}\ln\left(\frac{H_{\rm inf}M_{p}}{(1~{\rm GeV})^2}\right) -\frac{1}{3}\ln\left(\frac{T_{\rm rh}}{1~{\rm GeV}}\right)
\label{25}
\end{equation}
where
\begin{equation}
N_{\rm inf}\simeq 34+\ln\left(\frac{T_{\rm rh}}{100~{\rm GeV}}\right)~~\textrm{and}~~N_{\rm rh} \simeq \frac{2}{3}\ln\left(\frac{H_{\rm inf}M_{p}}{T_{\rm rh}^2}\right)-1.89
\label{26}
\end{equation}
In Eq. (\ref{24}) the parameter $ \kappa_X $ is taken to be large in order to increase the asymmetry. Also, the infrared cut-off $\mu$ must be sufficiently small. Here we assume minimal box cut-off \cite{Lyth:2007jh}, $\mu=H_0\approx 10^{-42}$ GeV, which accounts for all the modes which are within the present Hubble horizon.   
 
A similar relation to Eq. (\ref{24}) can be derived for the mixed anomalies. In particular, these can be present in the case of gauged baryon number ($ U(1)_B $), as considered in \cite{Barrie:2014waa}. In the case of electroweak and hypercharge mixed anomalies the extra contribution to the total $X$ charge asymmetry is,
\begin{equation}
\eta_X^{\textrm{mixed}}=\frac{n_X^{\textrm{mixed}}}{s}\approx   (|{\cal A}_{2}|^{5/2}g_{1}^{5} +3|{\cal A}_{3}|^{5/2}g_{2}^{5}) \frac{15 }{4\pi^{13}g_{*}} \Gamma\left(\frac{3}{4}\right)^4 e^{-3 N_{\rm e}}\left(\frac{\kappa}{ \mu T_{\rm rh}^2}\right)^\frac{3}{2}~.
\label{27}
\end{equation} 
where $ \kappa=\frac{\phi'_0H_{\rm inf}^2}{ f_X} $, and we have assumed that they have the same IR cut-off $\mu$. The choice of IR cut-off will be discussed further in the next section.

In the derivation of the asymmetry parameters above, Eq. (\ref{24}) and (\ref{27}), we have assumed that the only non-negligible contribution to the generated charge is produced during the inflationary epoch. The conditions for the mechanism considered here may still be active during the radiation epoch, but the overall effect will be negligible as the push out of equilibrium is considered to be too small in later epochs; hence the total $ X $ charge is assumed to be conserved once inflation ends. One exception to this is the possibility of sphaleron redistribution which will violate both $ B $ and $ L $ equally. The mutual dilution of the charge and the entropy densities, after reheating, ensures there is no further dilution of the asymmetry parameter. No additional washout processes have been considered in the above derivation. 

In the following section we utilise the known properties of sphaleron transitions to determine the distribution of $X$ charges amongst fermionic species after the electroweak phase transition, if the reheating temperature is greater than the critical temperature ($ T_{c}\sim 100 $). How the sphaleron processes redistribute the $X$ charge is dependent on the specific model being considered: the type of charge gauged, the associated anomalies, and the properties of the new fermion/s introduced.

\section{Computing $\rho_{\rm DM}/\rho_{B}$ and $ \eta_{B} $ \label{Sec5}}
Now that the $ X $ charge asymmetry parameter has been calculated we can derive the predicted dark to luminous matter mass density ratio and the baryon asymmetry parameter. The generated $ X $ charge density can be decomposed into Standard Model and dark matter components as follows,
\begin{equation}
n_{X} = n_{X}^{SM} + n_{D} 
\label{densities} 
\end{equation}

The Standard Model component will have an associated $ B-L $ charge which will be reprocessed by the action of sphaleron transitions, at the electroweak phase transition ($ T_{c}\sim 100 $ GeV), into a known fermionic distribution. The dark matter candidate considered here will be unaffected by the sphaleron transitions as it is a singlet under the electroweak interactions, but this does not have to be the case. 
%It is assumed that $ T_{\rm rh} > T_{c} $ due to the strict constraint that was derived in our previous work \cite{Barrie:2014waa}, hence this redistribution will be expected to occur.

After the electroweak phase transition the $ B-L $ charge will be distributed between $ B $ and $ L $ charges as follows: $ (B-L)_{SM}= \frac{79}{28}B$ and $(B-L)_{SM}= -\frac{79}{51}L $. We require that the resultant Standard Model baryon number asymmetry is consistent with that which is observed Eq. (\ref{2}). From this we obtain the following relation,
\begin{equation}
\eta_B =\epsilon(\eta_X^{\textrm{mixed}-SM }+\eta_X^{\textrm{unmixed}-SM })
\label{asym}
\end{equation}
where $ \epsilon $ is a step function defined by:
\begin{equation}
\epsilon:=\epsilon(T_{\rm rh})=
\begin{cases}
\frac{28}{79} & T_{\rm rh}>T_{\rm c} \\
1 & T_{\rm rh}<T_{\rm c}
\end{cases}
\end{equation}

Henceforth we will assume that the mixed anomalies only contribute to the standard model sector $ \eta_X^{\textrm{mixed}}=\eta_X^{\textrm{mixed}-SM } $, as our dark matter candidate is sterile under the standard model gauge groups; although this does not have to be the case.

It is assumed that the $ X $ charge density generated is initially uniformly distributed between each of the applicable fermion degrees of freedom:
\begin{align}
& \eta_X^{SM} =\eta_X^{\textrm{mixed}}+ \frac{\sum_{i} N^{i}_{SM}|q^{i}_{SM}|}{\sum_{i} N^{i}_{SM}|q^{i}_{SM}|+ \sum_{i} N^{i}_{D}|q^{i}_{D}|}\eta_X^{\textrm{unmixed} } \\ & \eta_D=\frac{\sum_{i} N^{i}_{D}|q^{i}_{D}|}{\sum_{i} N^{i}_{SM}|q^{i}_{SM}|+ \sum_{i} N^{i}_{D}|q^{i}_{D}|}\eta_X^{\textrm{unmixed}}
\label{distrib} 
\end{align}

where the index $ i $ corresponds to the particle species, $ N_{i} $ is the corresponding number of degrees of freedom, and $ q_{i} $ is the associated $ X $ charge. Therefore, the baryon asymmetry parameter defined above is given by,
\begin{equation}
\eta_B =\epsilon\left(\eta_X^{\textrm{mixed}}+ \frac{\sum_{i} N^{i}_{SM}|q^{i}_{SM}|}{\sum_{i} N^{i}_{SM}|q^{i}_{SM}|+ \sum_{i} N^{i}_{D}|q^{i}_{D}|}\eta_X^{\textrm{unmixed}}\right)
\label{asym2} 
\end{equation}

The predicted dark matter to luminous matter mass density ratio is given by,

\begin{equation}
\frac{\rho_{D}}{\rho_{B}}=\frac{m_{\psi}}{m_{B}}\frac{q_{B}\eta_{D}}{q_{\psi}\eta_{B}}=\frac{\eta_{D}}{q_{\psi}\eta_{B}}\left(\frac{m_{\psi}}{1~ \textrm{GeV}}\right)
\end{equation}
where we have assumed $ m_{B}=1 $ GeV and $ q_{B}=1 $. Now to consider this framework in the two scenarios introduced in Sec. \ref{Sec2}: gauged $ B-L $ and gauged $ B $ number, each including a sterile fermion charged under the given group.

\subsection*{Case 1: $U(1)_{B-L}$ and a sterile fermion}

In this scenario we must sum over all the Standard model fermions, $ \sum_{i} N^{i}_{SM}|q^{i}_{SM}|=21 $, assuming no RH neutrinos have been added. Only the unmixed cubic anomaly contributes to the $ B-L $ charge generation. Using these facts and requiring that Eq. (\ref{asym2}) is consistent with Eq. (\ref{2}) gives the following dark matter to luminous matter mass density ratio:
 \begin{equation}
  \frac{\rho_{D}}{\rho_{B}}\approx \frac{1}{21\epsilon} \left(\frac{m_{\psi}}{1 ~\textrm{GeV}}\right) ~~ \Rightarrow ~~ m_{\psi} \approx 116\epsilon ~\textrm{GeV}
  \end{equation}
  where we have chosen $ N_{\psi}=1 $ and also require that Eq. (\ref{1}) is satisfied.
 
It is found that the dark matter candidate $ \psi $ must have a mass $ m_{\psi}\approx 41$ GeV, or $ m_{\psi}\approx 116$ GeV; for $ T_{\rm rh}>T_c $ and $ T_{\rm rh}<T_c $ respectively. Interestingly, this ratio is found to be only dependent the mass of the associated dark fermion, and independent of the $ B-L $ charge of the dark matter candidate. Although it must be noted that this relation assumes the correct baryon asymmetry parameter is generated, and hence the parameters of the model are constrained.

The required replication of the observed baryon asymmetry results in the following condition on the model parameters, 
\begin{align}
 \eta_{B} & \approx \epsilon|{\cal A}_{1}|^{5/2} \frac{101}{21+|q_{\psi}|} \frac{1}{\pi^{13}g_{*}}\frac{m_{X}^5}{f_{X}^{5}} e^{-3 N_{\rm e}}\left(\frac{\kappa}{ \mu T_{\rm rh}^2}\right)^\frac{3}{2} \\ &
 \approx  3.5\times 10^{-18} ~\textrm{GeV}^{-1/2}~ \epsilon\frac{|{\cal A}_{1}|^{5/2}}{21+|q_{\psi}|}\frac{m_{X}^5}{f_{X}^{5}} \frac{H_{\rm inf}}{T_{\rm rh}^2}\left(\frac{\phi'_0}{ f_{X}}\right)^\frac{3}{2}
 \end{align}
It is found that this can satisfy Eq. (\ref{2}) for a wide range of parameter values.

\subsection*{Case 2: $U(1)_{B}$ and a sterile baryon}
If we now consider a gauged baryon number extension to the Standard Model we must sum over all of the baryonic degrees of freedom, $ \sum_{i} N^{i}_{SM}|q^{i}_{SM}|=12 $. In this scenario the contributions of the mixed anomalies $SU(2)^2\times U(1)_{B}$ and $U(1)_Y^2\times U(1)_{B}$ must be included, which generate a net charge only in the form of luminous matter. Hence we find that the dark matter to luminous matter mass density ratio is given by,
\begin{align}
\frac{\rho_{D}}{\rho_{B}} & =\frac{1}{\epsilon}\frac{N_{\psi}}{12+N_{\psi}|q_{\psi}|}\frac{ |\mathcal{A}_{1}|^{5/2}}{ \frac{12 |\mathcal{A}_{1}|^{5/2}}{N_{\psi}|q_{\psi}| +12}\frac{m_{X}^5}{f_{X}^{5}} +|\mathcal{A}_{2}|^{5/2}g^5_{1}+3|\mathcal{A}_{3}|^{5/2}g^5_{2} }\frac{m_{X}^5}{f_{X}^{5}}\left(\frac{m_{\psi}}{1~ \textrm{GeV}}\right) \\  & \approx \frac{1}{\epsilon}\frac{|q_{\psi}|^{15/2}}{12 |q_{\psi}|^{15/2}\frac{m_{X}^5}{f_{X}^{5}} +|q_{\psi}| +12}\frac{m_{X}^5}{f_{X}^{5}}\left(\frac{m_{\psi}}{1~ \textrm{GeV}}\right)
\label{barratio}
\end{align}
where  $ m_{B}=1 $ GeV and $ q_{B}=1 $ have been set. In the second line we have taken $ g_1^2 \simeq\frac{4\pi}{60}$and $ g_2^2 \simeq\frac{4\pi}{29}$, and used the anomaly values given in Sec. \ref{Sec2}: ${\cal A}_{2}=3/2$ and ${\cal A}_{3}=-3/2$. Upon rearranging and requiring Eq. (\ref{1}) and (\ref{2}), we find the following expression for the mass of the dark matter candidate,
\begin{equation}
m_{\psi}\approx \epsilon\frac{f_{X}^5}{m_{X}^{5}}\frac{11(12 |q_{\psi}|^{15/2}\frac{m_{X}^5}{f_{X}^{5}} +|q_{\psi}| +12)}{2 |q_{\psi}|^{15/2} } ~\textrm{GeV}
\label{barmass}
\end{equation}

It is expected that the required dark matter mass would be greater than that found in the previous case due to the additional contributions to the luminous sector from the mixed anomalies. For example, if we let $ |q_{\psi}|=1 $ and $ g_{X}^2=\frac{m_{X}}{f_{X}}\sim 0.01 $ then $ m_{\psi}\approx 2.5\times 10^6 $ GeV or $ m_{\psi}\approx 7.2\times 10^6 $ GeV, for $ T_{\rm rh}<T_{c} $ and $ T_{\rm rh}>T_{c} $ respectively. Once again in the derivations of Eq. (\ref{barratio}) and (\ref{barmass}) we have assumed the satisfaction of Eq. (\ref{2}).

The associated constraint imposed by the observed baryon asymmetry is given by
 \begin{align}
 \eta_{B} & \approx \epsilon\frac{\mathcal{A} }{\pi^{13}g_{*}}  e^{-3 N_{\rm e}}\left(\frac{\kappa}{ \mu T_{\rm rh}^2}\right)^\frac{3}{2} \\ & \approx  3\times 10^{-19} ~\textrm{GeV}^{-1/2}~ \epsilon\mathcal{A} \frac{H_{\rm inf}}{T_{\rm rh}^2}\left(\frac{\phi'_0}{ f_{X}}\right)^\frac{3}{2} 
 \end{align}
where $ \mathcal{A}=\left( \frac{12 |q_{\psi}|^{15/2}}{|q_{\psi}| +12}\frac{m_{X}^5}{f_{X}^{5}} +|\mathcal{A}_{2}|^{5/2}g^5_{1}+3|\mathcal{A}_{3}|^{5/2}g^5_{2} \right) $.

Similarly to the previous case, this can be satisfied for a wide range of parameter values. Although due to the dependence of $ m_{\psi} $ on both $ g_{X} $ we will have extra constraints on the inflationary expansion rate $ H_{\rm inf} $ because we have assumed $m_{\psi}\ll H_{\rm inf} $ throughout our derivations. This approximately translates to requiring that $ g_{X}^2=\frac{m_{X}^2}{f_{X}^2} \gtrsim 10^{-5} $, for $ q_{\psi}= 1 $ and is relaxed for larger charges.

\section{Conclusion}

In this paper we have investigated a model for simultaneous generation of luminous and dark matter during the inflationary epoch through the introduction of an anomalous gauge interaction and sterile fermion to the Standard Model.
It has been found that this scenario for cogenesis can be successfully reproduce observations for the two possible cases considered: gauged $ B $ and gauged $ B-L $ charge.
The general mechanism for cogenesis developed here could be applied to more complex models involving other or extra anomalous gauge symmetries and additional sterile/non-sterile fermionic states. It is possible that these additions could lead to a lessening of the parameter constraints imposed by the observed matter-antimatter asymmetry through extra contributions to the luminous matter generation. This could also alter the required mass of the dark matter candidate. Further study of these possibilities and the associated collider phenomenology is of interest.

\paragraph{Acknowledgment.} This work was partially supported by the Australian Research Council. 

\appendix
\section{Appendix: Further details of calculations}
\label{A}

Matching the solutions $ G_{\pm} $ (Eqs. (\ref{G+}) and (\ref{G-})) to the $\mathcal{CP}$-invariant solutions in Eq. (\ref{16}) in the limit $ |\tau|\rightarrow 0 $ allows the determination of the coefficients $ C_{1} $ and $ C_{3} $. 
\begin{equation}
C_{1}=\frac{\sqrt{\pi}\Gamma(\frac{1+\nu}{2}-\frac{\Omega_{k}}{4})}{2^{\frac{1}{2}(3-\nu)}}\left(\frac{k}{\Omega_{k}}\right)^{\nu} e^{i\frac{\pi}{2}(\frac{1}{2}-\nu)}
\end{equation}

and

\begin{equation}
C_{3}=\frac{\Gamma(\frac{1+\nu}{2}-\frac{i\Omega_{k}}{4})}{2^{\frac{1}{2}(3-\nu)}\sqrt{\pi}}\left(\frac{k}{\Omega_{k}}\right)^{\nu} e^{-i\frac{\pi}{4}}
\end{equation}

We then use the Wronskian normalisation $ \mathcal{W}(G_{\pm},G_{\pm}^*)=i $ to find $ C_{2} $ and $ C_{4} $. 

\begin{equation}
C_{2}=\frac{k^{\nu}}{2^{\frac{1}{2}(1+\nu)}\sqrt{\pi}}e^{-i\frac{\pi}{2}(\frac{1}{2}+\nu)}
\end{equation}

and

\begin{equation}
C_{4}=\frac{\sqrt{\pi}k^{\nu}e^{i\frac{\pi}{2}(\frac{1}{2}+\nu)}}{2^{\frac{1}{2}(3-\nu)}\Gamma(1+\nu)}\frac{\Gamma\left(\frac{1+\nu}{2}+\frac{i\Omega_{k}}{4}\right)}{\Gamma\left(\frac{1-\nu}{2}+\frac{i\Omega_{k}}{4}\right)}\textrm{Im} \left(\frac{\Gamma\left(\frac{1+\nu}{2}+\frac{i\Omega_{k}}{4}\right)}{\Gamma\left(\frac{1-\nu}{2}+\frac{i\Omega_{k}}{4}\right)} \right)^{-1}\left(1+e^{\frac{\pi\Omega_{k}}{4}}\frac{|\Gamma(\frac{1+\nu}{2}-\frac{i\Omega_{k}}{4})|^2}{\pi 2^{1-2\nu} \Omega_{k}^\nu}\right)
\end{equation}

In the calculation of these coefficients we utilised the following properties of Hypergeometric functions \cite{NIST-DLMF}:
In the limit $ z\rightarrow 0 $,
\begin{equation}
M(a,b,z)=1+O(z)
\end{equation}

\begin{equation}
U(a,b,z)=\frac{\Gamma(b-1)}{\Gamma(a)}z^{1-b} + \frac{\Gamma(1-b)}{\Gamma(a-b+1)} O(z^{2-b}).
\end{equation}

and the known Wronskian relations,
\begin{equation}
\mathcal{W}\{M(a,b,z),U(a,b,z) \}=-z^{-b} e^{-z}\Gamma(a)\Gamma(b)
\end{equation}

\begin{equation}
\mathcal{W}\{M(a,b,z),e^{z}U(b-a,b,e^{\pm i \pi}z) \}=e^{\mp b i \pi}z^{-b} e^{z}\Gamma(b-a)\Gamma(b)
\end{equation}

\begin{equation}
\mathcal{W}\{U(a,b,z),e^{z}U(b-a,b,e^{\pm i \pi}z) \}=e^{\mp (a-b) i \pi}z^{-b} e^{z}
\end{equation}

\end{document}